\documentclass{article}

\usepackage{a4}

\scrollmode
\usepackage{amsmath}
\usepackage{amsfonts}
\usepackage{amssymb}
\usepackage{latexsym}
\usepackage{stmaryrd}
\usepackage{array}
\usepackage{exscale}
\newcommand{\nc}{\newcommand}

\newcommand{\ve}{\varepsilon}

\newcommand{\sse}{\subseteq}

\newcommand{\spe}{\supseteq}
\newcommand{\fa}{\forall}

\newcommand{\mr}{\mathrm}
\newcommand{\mc}{\mathcal}
\newcommand{\mf}{\mathfrak}

\newcommand{\DMO}{\DeclareMathOperator}
\newcommand{\DST}{\displaystyle}

\newcommand{\RR}{\mathbb{R}}

\mathchardef\breakingcomma\mathcode`\,
{\catcode`,=\active
  \gdef,{\breakingcomma\discretionary{}{}{}}
}

\usepackage{listings}
\newcommand{\inl}[1]{\lstinline$#1$}
\newcommand{\set}[1]{\{ #1 \}}

\nc{\simlvi}[1]{\!\sim_{#1}}
\nc{\apprel}[3]{{#1}(#2)_{(#3)}} 
\nc{\cmpli}[1]{\complement^1_{#1}} 
\nc{\cmplzi}[1]{\complement^0_{#1}} 
\nc{\cmplzoi}[1]{\complement^*_{#1}} 

\nc{\zf}{\mr{ZF}}
\nc{\zfmf}{\zf^0} 
\nc{\zfc}{\mr{ZFC}}
\nc{\zfcmf}{\zfc^0} 
\nc{\bst}{\mr{BST}} 
\providecommand{\abs}[1]{\lvert #1 \rvert} 
\providecommand{\norm}[1]{\lVert #1 \rVert} 
\makeatletter
\DeclareRobustCommand{\genericinterval}[2]{%
  \@ifstar{\genericinterval@star{#1}{#2}}{\genericinterval@nostar{#1}{#2}}}
\newcommand{\genericinterval@star}[4]{\mathopen{}\mathclose{\left#1#3,#4\right#2}}
\newcommand{\genericinterval@nostar}[4]{\mathopen{#1}#3,#4\mathclose{#2}}

\makeatother
\nc{\untit}[2]{{#1}^{#2 \downarrow}} 
\nc{\obit}[2]{{#1}^{#2 \uparrow}} 
\nc{\inzEKi}[1]{\mc{I}^{\mr{V}}_{#1}}

\nc{\inzKEi}[1]{\mc{I}^{\mr{E}}_{#1}}

\nc{\adjEi}[1]{\mc{A}^{\mr{V}}_{#1}}

\nc{\BD}[1]{{#1}\text{-}\mr{BD}}

\nc{\konv}[2]{{#1}[{#2}]} 
\nc{\actpres}[1]{\Phi_{#1}} 
\nc{\Prim}{\mc{PR}} 

\nc{\sselr}{\sse^{\mapsto}}
\nc{\sserl}{\sse^{\mapsfrom}}
\nc{\spelr}{\spe^{\mapsto}}
\nc{\sperl}{\spe^{\mapsfrom}}
\nc{\ball}[1]{\mr{B}^{#1}} 
\nc{\oball}[1]{\breve{\mr{B}}^{#1}} 
\nc{\pball}[1]{\dot{\mr{B}}^{#1}} 
\nc{\prr}[1]{\dot{\RR}^{#1}} 
\nc{\sph}[1]{\mr{S}^{#1}} 
\nc{\ssim}[1]{s\sigma_{#1}} 
\nc{\koerper}[1]{\norm{#1}}
\nc{\Ccovdim}{\mc{CD}}
\nc{\Cinddim}{\mc{SID}}

\nc{\CInddim}{\mc{LID}}

\DeclareMathOperator{\diffop}{D} 
\DeclareMathOperator*{\diffoplimit}{D} 
\nc{\diffopc}[1]{\sideset{_{#1}}{}\diffoplimit} 
\nc{\diffopp}[1]{\diffop_{#1}} 
\nc{\diffopcp}[2]{\sideset{_{#2}}{_{#1}}\diffoplimit} 
\nc{\meanH}[2]{\mf{M}_{#1,#2}} 
\nc{\emean}[2]{\mf{M}_{\exp_{#1},#2}} 
\DeclareMathOperator{\mor}{Mor}
\DeclareMathOperator{\Hom}{Hom} 
\nc{\autoerw}[1]{\mr{Aut}^{#1}} 
\nc{\komma}[2]{(#1 \downarrow #2)} 
\nc{\Kmat}{\mf{MAT}} 
\nc{\Khmat}{\mf{HMAT}} 
\nc{\homfun}[1]{\mor_{#1}(-_1,-_2)} 
\nc{\homfunae}[1]{\mor_{#1}(-_1)} 
\nc{\homfunbe}[1]{\mor_{#1}(-_2)} 
\nc{\homfunxy}[3]{\mor_{#1}(#2(-_1), #3(-_2))}
\nc{\homfunx}[2]{\mor_{#1}(#2(-_1), -_2)}
\nc{\homfuny}[2]{\mor_{#1}(-_1, #2(-_2))}
\nc{\homfuna}[2]{\mor_{#1}(#2, -)} 
\nc{\homfunb}[2]{\mor_{#1}(-, #2)} 
\nc{\hhomfuna}[2]{\Hom_{#1}(#2, -)} 
\nc{\hhomfunb}[2]{\Hom_{#1}(-, #2)} 

\newcommand{\Cls}{\mc{CLS}}

\newcommand{\Musat}{\mc{M\hspace{0.8pt}U}} 
\newcommand{\Musati}[1]{\Musat_{\!#1}} 
\newcommand{\Smusat}{\mc{S}\Musat} 
\newcommand{\Smusati}[1]{\Smusat_{\!#1}}

\nc{\Clsoo}{\Cls^{1,1}} 

\DeclareMathOperator{\var}{var}

\DMO{\dos}{ds} 
\DMO{\mdos}{mds} 
\newcommand{\Clash}{\mc{HIT}} 

\newcommand{\Uclash}{\mc{U}\Clash} 
\newcommand{\Uclashi}[1]{\Uclash_{\!\!#1}}

\DMO{\premr}{ax} 
\DMO{\concr}{C} 
\DMO{\allcr}{cl} 

\DMO{\thardness}{thd} 
\DMO{\phardness}{phd} 
\DMO{\whardness}{awid} 
\DMO{\dep}{dep} 
\DMO{\hts}{hs} 
\DMO{\semspace}{css} 
\DMO{\resspace}{crs} 
\DMO{\treespace}{cts} 
\nc{\bth}[1]{\langle{#1}\rangle} 
\DMO{\rsub}{r_S} 
\DMO{\rk}{r} 
\DMO{\ro}{\rk_1} 
\DMO{\rki}{\rk_{\infty}} 
\DMO{\rpl}{r^{pl}} 
\DMO{\ropl}{\rk_1^{pl}} 
\nc{\rslur}{\xrightarrow{\text{SLUR}}} 
\nc{\rslurs}{\rslur_{\!*}} 
\DMO{\slur}{slur} 
\nc{\Slur}{\mc{SLUR}} 
\nc{\rkslur}[1]{\xrightarrow{\text{SLUR}_{#1}}} 
\nc{\rkslurs}[1]{\rkslur{#1}_{\!*}} 
\nc{\Altsluri}[1]{\Slur(#1)}
\nc{\Altslurstari}[1]{\Slur\text{\textasteriskcentered}(#1)}
\nc{\Canoni}[1]{\mr{CANON}(#1)}
\nc{\rkslurstar}[1]{\xrightarrow{\text{SLUR\textasteriskcentered}#1}} 
\nc{\rkslursstar}[1]{\rkslurstar{#1}_{\!*}} 
\DMO{\slurstar}{\slur\!\text{\textasteriskcentered}}
\nc{\Urefc}{\mc{UC}}
\nc{\Propc}{\mc{PC}}
\nc{\Wrefc}{\mc{WC}} 
\DeclareMathOperator{\vdeg}{vd} 
\DeclareMathOperator{\minvdeg}{\mu\!\vdeg} 
\DMO{\varmvd}{\var_{\minvdeg}} 
\DMO{\nfc}{fc} 
\DMO{\maxnfc}{\nu\!\nfc} 
\nc{\Dt}[1]{\mc{F}_{#1}} 

\nc{\svbf}{\mc{VB}} 
\nc{\svbfs}{\mc{VB}^*} 
\DMO{\potp}{pp} 
\DMO{\potprec}{NM} 
\DMO{\minnonmer}{VDM} 
\DMO{\minnonmerh}{VDH} 
\DMO{\maxsmar}{FCM} 
\DMO{\maxsmarh}{FCH} 
\DMO{\varsing}{\var_s} 
\DMO{\varosing}{\var_{1s}} 
\DMO{\varnosing}{\var_{\neg1s}} 
\DMO{\nsv}{\mathit{n}_s} 
\DMO{\nosv}{\mathit{n}_{1s}}
\DMO{\nnosv}{\mathit{n}_{\neg1s}}
\nc{\Musatns}{\Musat'} 
\nc{\Musatnsi}[1]{\Musati{#1}'}
\nc{\Smusatns}{\Smusat'} 
\nc{\Smusatnsi}[1]{\Smusati{#1}'}
\nc{\Uclashns}{\Uclash'} 
\nc{\Uclashnsi}[1]{\Uclashi{#1}'}
\nc{\tsdp}{\xrightarrow{\text{sDP}}}
\nc{\tsdps}{\tsdp_{\!*}}
\nc{\tosdp}{\xrightarrow{\text{1sDP}}}
\nc{\tosdps}{\tosdp_{\!*}}
\DMO{\sdp}{sDP} 
\DMO{\osdp}{sDP_1} 
\nc{\cflmusat}{\mc{CF}\Musat} 
\nc{\cflmusati}[1]{\mc{CF}\Musati{#1}}
\nc{\cflimusat}{\mc{CFI}\Musat} 
\DMO{\sNF}{sNF} 
\DMO{\eqp}{eqp} 
\DMO{\sgp}{sp} 
\DMO{\singind}{si} 
\DMO{\osingind}{si_1} 
\DMO{\shyp}{svh} 
\DMO{\sdph}{ssh} 
\DMO{\msdph}{mss} 
\DMO{\osdph}{ssh_1} 
\DMO{\mosdph}{mss_1} 
\DMO{\mps}{mps} 
\DMO{\purec}{puc} 
\DMO{\doping}{D}
\nc{\glue}[4]{\operatorname{glue}((#1,#2), (#3,#4))} 
\nc{\gluea}[3]{#1 \mathbin{\boxplus}_{#3} #2} 
\DMO{\saturate}{S}
\DMO{\marginalise}{M}
\DMO{\frl}{fl} 
\nc{\Con}{\mr{Con}}
\nc{\Log}{\mr{Log}}
\nc{\Lin}{\mr{Lin}}
\nc{\Pol}{\mr{Pol}}
\nc{\ExL}{\mr{ExL}}
\nc{\ExP}{\mr{ExP}}
\nc{\CTime}{\mr{CTime}}
\nc{\CSpace}{\mr{CSpace}}
\nc{\LTime}{\mr{LTime}}
\nc{\LSpace}{\mr{L}}
\nc{\NLSpace}{\mr{NL}}
\nc{\LinTime}{\mr{LinTime}}
\nc{\LinSpace}{\mr{LinSpace}}
\nc{\PTime}{\mr{P}}
\nc{\PSpace}{\mr{PSpace}}
\nc{\Np}{\mr{NP}}
\nc{\Conp}{\text{coNP}}
\nc{\NPSpace}{\mr{NPSpace}}
\nc{\CoNPSpace}{\mr{coNPSpace}}
\nc{\ELTime}{\mr{ELTime}}
\nc{\ELSpace}{\mr{ELSpace}}
\nc{\EPTime}{\mr{EPTime}}
\nc{\EPSpace}{\mr{EPSpace}}
\nc{\NEPTime}{\mr{NEPTime}}
\nc{\polydelta}[1]{\Delta_{#1}^{\mr P}}
\nc{\polypi}[1]{\Pi_{#1}^{\mr P}}
\nc{\polysigma}[1]{\Sigma_{#1}^{\mr P}}
\nc{\Ph}{\mr{PH}}

\nc{\Dp}{D^P}
\nc{\PllC}[2]{{\text{$\mr{PT}$/$\mr{WK}$}(#1, #2)}} 
\nc{\Nc}{\mr{NC}}
\nc{\Nci}[1]{\Nc^{#1}}
\nc{\Ac}{\mr{AC}}
\nc{\Aci}[1]{\Ac^{#1}}
\nc{\pmodpoly}{P / \mathrm{poly}}
\nc{\Wh}[1]{\mr{W}[#1]} 
\nc{\Rl}{\mr{RL}}
\nc{\coRl}{\mr{coRL}}
\nc{\Rp}{\mr{RP}}
\nc{\coRp}{\mr{coRP}}
\nc{\Zpp}{\mr{ZPP}}
\nc{\Bpp}{\mr{BPP}}
\nc{\Pp}{\mr{PP}}
\nc{\Reach}{\mr{STCON}} 
\nc{\Undreach}{\mr{USTCON}} 
\nc{\Pcol}[2]{\mr{COL}(#1,#2)} 
\nc{\Pscol}[2]{\mr{SCOL}(#1,#2)} 
\nc{\Psorcol}[2]{\mr{SORCOL}(#1,#2)} 
\DMO{\slp}{slp}
\nc{\Mss}{\mr{MSS}}
\nc{\Key}{\mr{KEY}}
\nc{\Keyi}[1]{\Key_{\!#1}}
\nc{\Nbmss}{N_{\mr{bm}}} 
\nc{\Nbkey}{N_{\mr{bk}}} 
\nc{\Rnb}{N_{\mr{b}}}
\nc{\Rnk}{N_{\mr{k}}}
\nc{\Rnr}{N_{\mr{r}}}

\nc{\Byte}{\mr{B}[8]}
\nc{\QByte}{\mr{B}[4,8]}
\nc{\KByte}{\mc{B}} 
\nc{\RQByte}{\mc{QB}} 

\nc{\ramz}[3]{\mr{ram}_{#1}^{#2}(#3)} 
\nc{\waez}[2]{\mr{vdw}_{#1}(#2)} 
\nc{\gtz}[2]{\mr{grt}_{#1}(#2)} 
\nc{\pdwaez}[2]{\mr{vdw}_{#1}^{\mr{pd}}(#2)} 
\nc{\absfeh}[1]{\delta_{#1}} 
\nc{\relfeh}[1]{\ve_{#1}} 
\usepackage{theorem} 
\usepackage[breaklinks]{hyperref}
\theorembodyfont{\rmfamily}

\theorembodyfont{}
\newcounter{dDef} 

\newcounter{dLem} 

\newcounter{dThm} 

\newcounter{dPro} 

\newcounter{Beispielzaehler}

\nc{\bm}{\boldmath}
\nc{\bmm}[1]{\mbox{\bm$\DST #1$}}
\nc{\mi}[1]{\bmm{\mathrm{(#1):}} \quad}

\usepackage{enumerate}
\usepackage[active]{srcltx}
\usepackage[all]{xy}

\begin{document}

\title{Minimal unsatisfiability and deficiency:\\ recent developments}

\author{
  \href{http://cs.swan.ac.uk/~csoliver}{Oliver Kullmann}\\
  Computer Science Department\\
  Swansea University\\
  Swansea, SA2 8PP, UK
}

\maketitle

\begin{abstract}
  Starting with Aharoni and Linial \cite{AhLi86}, the \emph{deficiency} $\delta(F) = c(F) - n(F) \ge 1$ for minimally unsatisfiable clause-sets $F \in \Musat$, the difference of the number of clauses and the number of variables, is playing an important role in investigations into the structure of $\Musat$. In my talk\footnote{\url{http://cs.swan.ac.uk/~csoliver/papers.html\#BORDEAUX2016}} I want to give a high-level overview on recent developments.
\end{abstract}

$\Musat$ is the set of clause-sets $F$, which are unsatisfiable, while removal of any clause renders $F$ satisfiable; one can say that $F \in \Musat$ (``MU'') presents a ``single reason'' for unsatisfiability, while a general unsatisfiable $F$ may contain many $F' \sse F$ with $F' \in \Musat$ (``MUSs''), and thus has ``many reasons'' for unsatisfiability (see \cite{LPMMS2016MUS} for a recent work on finding MUSs).

Using $n(F)$ for the number of variables actually occurring in $F$ and $c(F) := \abs{F}$ for the number of clauses, the \emph{deficiency} is $\delta(F) := c(F) - n(F)$. There are many proofs of ``Tarsi's Lemma'' $\fa\, F \in \Musat : \delta(F) \ge 1$ (for a recent overview on them see the introduction of \cite{KullmannZhao2010Extremal}). By \cite{FKS00} we know that deficiency is a proper complexity parameter, each level $\Musati{\delta=k} := \set{F \in \Musat : \delta(F) = k}$ for $k=1,2,\dots$ polytime-decidable, with growing complexity. The level $k=1$ is well-known (\cite{DDK98}), the level $k=2$ also quite well (\cite{KleineBuening2000SubclassesMU,KullmannZhao2012ConfluenceJ}), while beyond that for general $\Musat$ still not much is known (though slowly this is changing). A general overview on $\Musat$ is given in \cite{Kullmann2007HandbuchMU}, while a recent extensive overview, also discussing the various connections to combinatorics, is the introduction of \cite{KullmannZhao2010Extremal}. Currently there are (at least) developments in the following areas concerning $\Musat$ as layered by deficiency:
\begin{itemize}
\item The \emph{minimum} degree of variables (minimum number of occurrences) is studied in \cite{KullmannZhao2011Bounds,KullmannZhao2010Extremal}, which seems an intricate problem, especially when looking for precise numbers. The maximum over all minimum degrees for deficiency $k$ is denoted by $\minnonmer(k)$.
\item Knowing $\minnonmer(1) = 2$ (i.e., for deficiency $1$ there must exist a variable occurring only twice), it is relatively easy to gain understanding of $\Musati{\delta=1}$, while $\minnonmer(2) = 4$ leaves some work for $\Musati{\delta=2}$, but is a good start. The next frontier is $\minnonmer(3) = 5$ --- so $F \in \Musati{\delta=3}$ has a variable occurring in one sign at most $3$ times, and in the other sign at most $2$ times, and by this information there is a chance to reconstruct $F$ from $\Musati{\delta\le 2}$.
\item A refined parameter is the number $\nfc(F)$ of \emph{full clauses} for $F \in \Musat$, the number of clauses $C \in F$ with $\abs{C} = n(F)$, where the maximum for deficiency $k$ is denoted by $\maxsmar(k)$; obviously we have $\maxsmar(k) \le \minnonmer(k)$. We have $\maxsmar(1) = 2$, $\maxsmar(2) = 4$ (exercise: find the easy examples), and $\maxsmar(3) = 4 = \minnonmer(3) - 1$ (\cite{KullmannZhao2010Extremal}; creating an example is a bit tougher now, and the upper bound needs some insight).
\item An important restriction for us is $\Uclash \subset \Musat$, the set of unsatisfiable \emph{hitting clause-sets}, that is, where every two different clauses have at least one clashing literal pair. Considering only such $F \in \Uclash$, the maximum of $\nfc(F)$ over $\Uclashi{\delta=k}$ is denoted by $\maxsmarh(k) \le \maxsmar(k)$. This parameter is studied in \cite{KullmannZhao2015FullClauses}, drawing new connections to number theory and the study of certain recursions, started in \cite[Page 145]{Hofstadter1979GEB}, and today called ``meta-Fibonacci recurrences'' (\cite{Conolly1989MetaFibonacci}).
\item The big theorem on the horizon, explaining the structure of $\Musat$, is a proof (and formulation) of the \emph{Finite Patterns Conjecture}, as discussed in the outlook of \cite{KullmannZhao2010Extremal}: for every deficiency $k$ we can describe the elements of $\Musati{\delta=k}$ via finitely many ``patterns'' (currently for $k \ge 3$ even the precise meaning of ``pattern'' is not known).
\item For $\Uclash$ this can be said precisely: the conjecture is that the number of variables of $F \in \Uclashi{\delta=k}$ for fixed $k$ is bounded, after elimination of ``singular'' variables, that is, variables occurring in one sign only once. And indeed we conjecture the maximum to be equal $4k - 5$ for $k \ge 2$. For $k=2$ this is known, while we established it for $k=3$ in \cite{KullmannZhao2016UHitSAT}. This (non-trivial) proof uses a variety of \emph{reductions}, the most basic one being \emph{singular DP-reduction}, DP-reduction (also called ``variable elimination'') applied to a singular variable.
\item The study of singular DP-reduction for $\Musat$ is the topic of \cite{KullmannZhao2012ConfluenceJ}, containing results related to \emph{confluence} (the result of the reduction is independent of the choices made during the (nondeterministic) reduction). The main result here is, that in general the number of variables in the result is unique, while for deficiency $2$ moreover the isomorphism type of the result is unique.
\item More generally the concept of \emph{clause-irreducibility} is developed in \cite{KullmannZhao2016UHitSAT} (with a forerunner in number theory; see \cite{Korec1984Covers,BergerFelzenbaumFraenkel1990Covers}). This surprisingly powerful concept says, that an unsatisfiable $F$ is clause-irreducible iff there is no $F' \subset F$ with $c(F') > 1$ such that $F'$ is logically equivalent to a single clause.
\end{itemize}

I want to motivate and give examples for the above developments in my talk. An interesting algorithmic problem is posed in \cite{KullmannZhao2010Extremal}:
\begin{enumerate}
\item A general upper bound on $\minnonmer(k)$ is developed, and it is shown that this bound is (precisely) sharp for \emph{lean clause-sets}, which are clause-sets without non-trivial autarkies (lean clause-sets were introduced in \cite{Ku00f}, but indeed already implicitly considered in \cite{KleeLadner1981WeakSAT} via ``weak satisfiability'').
\item As an aside, it is interesting to note here, that the bound is already sharp for \emph{variable-minimal unsatisfiable} clause-sets, as introduced in \cite{ChenDing2006VMU}, and further studied in \cite{KullmannZhao2010Extremal}, also correcting various errors from \cite{ChenDing2006VMU}.
\item So, if the bound is violated, then there must be an autarky!
\item We can indeed simulate the corresponding autarky reduction (as introduced in \cite{MoSp85}, removing the clauses satisfied by the autarky) in polynomial time, exploiting the refinement of deficiency by \emph{surplus}, as first studied in \cite{Szei2002FixedParam}.
\item But how to \emph{find} the autarky in polynomial time is an open problem. The underlying problem is to find for some polytime decidable class of satisfiable clause-sets a satisfying assignment in polynomial time.
\end{enumerate}

\bibliographystyle{plainurl}


\newcommand{\noopsort}[1]{}

\end{document}